\documentclass[aip,jap,superscriptaddress,noshowpacs,nobalancelastpage,preprint]{revtex4-1}
\usepackage{bm}
\usepackage{color}
\usepackage{textcase, natbib, url}
\usepackage[normalem]{ulem}
\usepackage[colorlinks=true, linkcolor=blue, urlcolor=black, citecolor=blue]{hyperref}
\usepackage{amsmath,amssymb}
\usepackage{mathtools}
\usepackage{tabularx}
\usepackage{lettrine}

\usepackage{natbib}

\begin{document}
	
\title{Efficient 3-dimensional photonic-plasmonic photo-conductive switches for picosecond THz pulses}

\author{Giorgos Georgiou}
\email[Corresponding author: ]{giorgos.georgiou@neel.cnrs.fr}
\affiliation{Univ. Savoie Mont-Blanc, CNRS, IMEP-LAHC, 73370 Le Bourget du Lac, France}
\affiliation{Univ. Grenoble Alpes, CNRS, Grenoble INP, Institut N\'eel, 38000 Grenoble, France}

\author{Cl\'{e}ment Geffroy}
\affiliation{Univ. Grenoble Alpes, CNRS, Grenoble INP, Institut N\'eel, 38000 Grenoble, France}

\author{Christopher B\"auerle}
\affiliation{Univ. Grenoble Alpes, CNRS, Grenoble INP, Institut N\'eel, 38000 Grenoble, France}

\author{Jean-Fran\c{c}ois Roux}
\affiliation{Univ. Savoie Mont-Blanc, CNRS, IMEP-LAHC, 73370 Le Bourget du Lac, France}

\begin{abstract}
The efficiency of photo-conductive switches, which continue to be used for the generation and detection of THz waves, has been overlooked for a long time. 
The so far ``optics-dominated'' devices are making their way through to new and emerging fields of research that require ultrafast picosecond voltage pulses, as well as to new applications where power efficiency is of uttermost importance.
To address the efficiency problems, in this article we present a novel photo-conductive switch that is based on a 3-dimensional design.
In contrast to conventional planar designs, our photo-conductive switch drastically enhances the overall efficiency by maximising the laser absorption within the device, while at the same time optimising the carrier collection efficiency at the electrodes. 
To maximise the optical absorption we take advantage of photonic and plasmonic modes that are excited in our device due to a periodic array of nanopillars, whereas the collection efficiency is optimised by converting each nanopillar into a single nano-photo-conductive switch. 
Our numerical calculations show a 50-fold increase in the overall generated current and a 5-fold bandwidth increase compared to traditional interdigitated planar photo-conductive switches. 
This opens up a wealth of new possibilities in quantum science and technology where efficient low power devices are indispensable.

\end{abstract}

\maketitle

%==========================================
%Introduction
%==========================================
Terahertz (THz) photo-conductive switches have been the workhorse for the fields of THz spectroscopy and ultrafast electronics. With their simple design and their straightforward implementation, photo-conductive switches based on interdigitated electrodes can generate and detect picosecond voltage pulses when illuminated by a femtosecond laser pulse. 

Since their development in the mid-80s\cite{auston1975, Mourou1981, Auston1984}, THz photo-conductive switches have been used primarily for spectroscopy in the THz frequency band (0.1-10 THz)\cite{Hangyo2005}. Additionally, due to their ability to generate picosecond voltage pulses, photo-conductive switches have also been used for probing ultrafast dynamics in semiconductors\cite{Ulbricht2011}, high frequency characterisation of electronic circuits\cite{Bieler15}, processing of radio-frequency (RF) signals\cite{Kuppam14} and high speed wireless communications\cite{Seeds2015}.
For a long period of time, research around THz photo-conductive switches was focused on improving their time response, namely from $15\ \rm{ps}$\cite{auston1975} to the sub-picosecond time scale\cite{Gupta1991}. A faster photo-conductive switch meant that higher THz frequencies could be achieved and therefore time-domain spectroscopy from the far- to the mid-infrared frequency band was possible. For these kind of photo-switches the generated THz power was not a major concern. This was because femtosecond lasers, which were used to power such photo-conductive switches, could provide plenty of optical power to operate any THz experiment with a very high signal-to-noise ratio. As a consequence, the ``classical'' interdigitated electrode photo-switch design was maintained despite its poor power conversion efficiency.

With an increasing demand for THz photo-conductive switches used for commercial purposes and for new fields of fundamental research\cite{wu2015, zhang2016, Yange2017,lin2018}, novel, more efficient designs had to be conceived. 
It was not until very recently that concepts from the fields of nano-photonics and plasmonics were incorporated into the designs of photo-conductive switches to enhance optical absorption and hence the overall conversion efficiency\cite{Lepeshov2017}.
Photo-conductive switches that incorporate plasmonic nano-resonators into their designs take advantage the large local field enhancements generated by the resonators, which in turn increases the optical absorption in the semiconductor\cite{heshmat2012, Park2012, Jarrahi2013, planken2014, moon2015, Burford2016, lepeshov2018, Bhattacharya2019}. 
Hybrid designs that combine distributed Bragg gratings (DBRs) with plasmonic resonators have also shown increased efficiency, since they can create cavities for the incoming laser light and trap it inside the semiconductor\cite{mitrofanov2015, Bashirpour2017, Mitrofanov2017, siday2019}. 
Although the majority of these studies have successfully addressed the problem of the laser absorption within the semiconductor, little work has been done on optimising the collection of the generated carriers by the contact electrodes.
Highly efficient photo-conductive devices are indeed necessary for quantum technology applications where the quantum devices are usually operated at cryogenic temperatures. 
For these new technologies it is important to control the quantum states on the picosecond timescale\cite{Bauerle2018}.

In this article, we present a novel architecture/design for THz photo-conductive switches which has a 3-dimensional photonic design. We demonstrate via numerical simulations that the proposed architecture has superior power conversion efficiency compared to the ``classical'' 2-dimensional planar devices.
By incorporating concepts and ideas from the fields of solar cells\cite{atwater2011,Liu2014,Liu2016}, we propose a design that can be operated on a very small bias voltage and with low optical power. 
The low power requirements are particularly attractive for experiments that can be sensitive to the heat generated by the laser or the photo-conductive switch, such as experiments performed at cryogenic temperatures\cite{zhong2008t,Parmentier2016}. 
Our 3D photo-conductive switch design aims at the complete absorption of the input laser power, with the majority of the optical power being absorbed by the semiconducting material and converted into photo-excited carriers. 
Furthermore, its unique vertical design can uniformly accelerate the generated carriers towards the collection electrodes.

In what follows, we will introduce the 3-dimensional design of the photo-conductive switch and how, with this architecture, we can optimise optical absorption through photonic and plasmonic resonant modes. 
This will be followed by a discussion on the electronic properties of the photo-conductive switch when a bias voltage is applied on the electrodes of the device. For completeness, we will numerically compare the performance of our photo-conductive switch with the conventional interdigitated electrode switches. 

%==========================================
% Main text
%==========================================
\subsection*{Photo-conductive switch design and photonic modelling}

\begin{figure}[t!]
%\begin{center}
\includegraphics[width=0.95\linewidth]{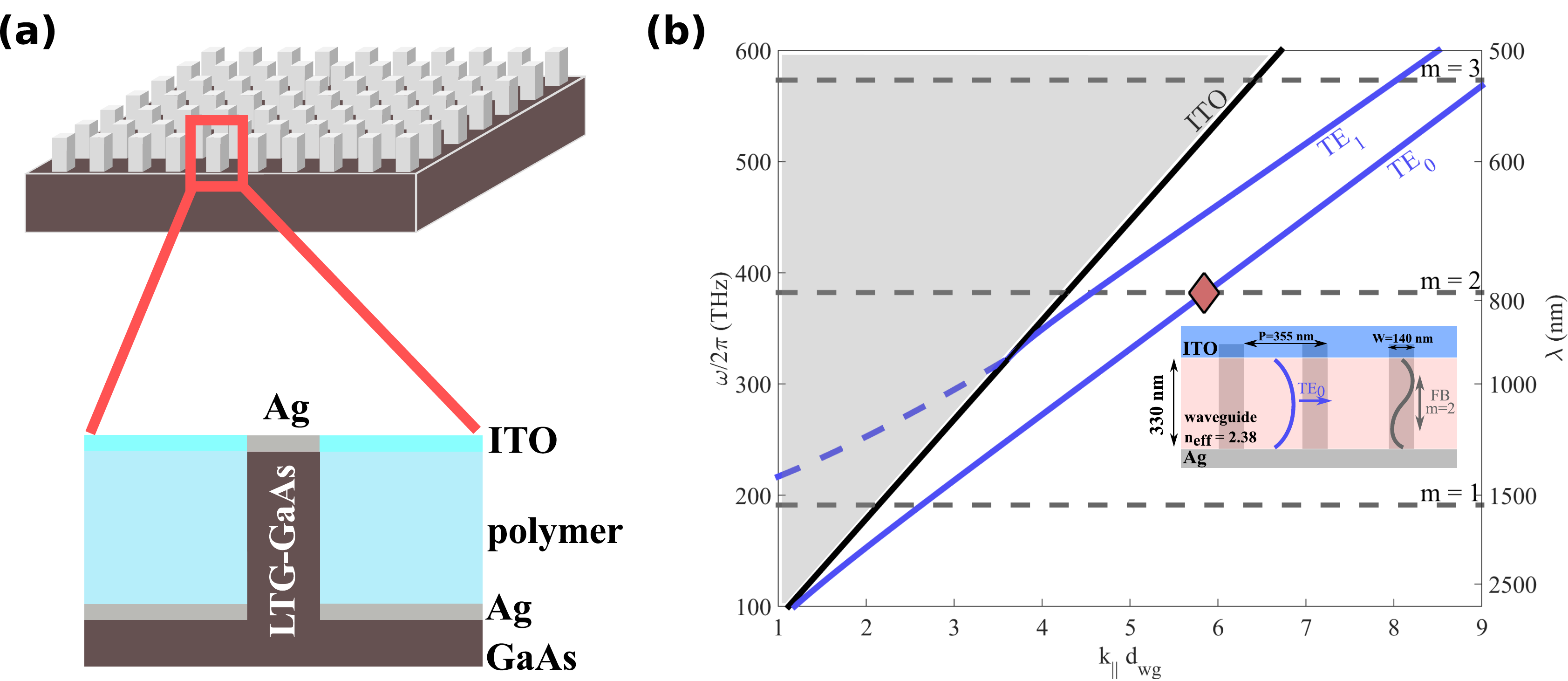}
\caption{\textbf{Vertical 3D photo-conductive switch.} \textbf{a}, Illustration of the 3-dimensional photo-conductive switch. The top image shows the array of nanopillars constituting the device and the bottom image shows a unit cell of the device. \textbf{b}, Dispersion relation for the 3-dimensional photo-conductive switch. The blue coloured lines illustrate the TE waveguide modes that can be supported by the device. These modes are bounded between the bottom metallic contact and the top transparent electrode (ITO). Modes excited on the left side of the ITO light cone (black line) are leaky (grey region). The large red diamond indicates the extra momentum added to an incident plane wave by the periodic arrangement of the nanopillars  $P=355\ \rm{nm}$. At $\lambda=780\ \rm{nm}$ we can couple the laser beam into the TE$_{0}$ waveguide mode. The grey dashed lines show the frequencies for which a Fabry-Perrot cavity mode can be excited in the vertical direction of our device. The $m=2$ mode coincides with the excitation of the TE$_{0}$ waveguide mode.}
\label{fig1:schematic_dispersion}
%\end{center}
\end{figure}

The proposed photo-conductive switch takes a leap forward from the traditional 2-dimensional planar designs. Figure \ref{fig1:schematic_dispersion}a shows an artistic illustration of our vertical 3-dimensional photo-conductive switch device. The device consists of several hundreds of vertical nanopillars that are connected in a parallel configuration. Each and every one of the nanopillars is an individual photo-conductive switch that can absorb light, accelerate and collect photo-generated carriers, resulting into photo-current. The nanopillars are made out of  a semiconducting material with its band gap energy being lower than the laser photon energy, typically low temperature grown (LTG-) GaAs, and are positioned in a periodic array. As will be discussed later, the periodic configuration facilitates an increased optical absorption in the semiconductor due to the excitation of photonic and plasmonic modes.
To accelerate and collect the photo-excited carriers, we position metallic electrodes covering completely the bottom surface of the device as well as the top part of each nanopillar. In between the nanopillars and up to the top metallic electrode, we place a transparent insulator with a low dielectric index ($n=1.58$) and high breakdown electric field. At the very top of the device, we connect all the top electrodes of all the nanopillars with a transparent electrode that allows the propagation of the incident laser beam through the interior of the device. Such a transparent contact can be an indium-tin-oxide (ITO) alloy with the right composition such that it has an increased optical transmission but low enough electrical resistance to avoid resistive losses.

%%%%%%%%%%%%%%%%%%%%%%%%% Fig. 1 %%%%%%%%%%%%%%
%%%%%%%%%%%%%%%%%%%%%%%%% Fig. 1 %%%%%%%%%%%%%%

\begin{figure}[t!]
%\begin{center}
\includegraphics[width=0.98\linewidth]{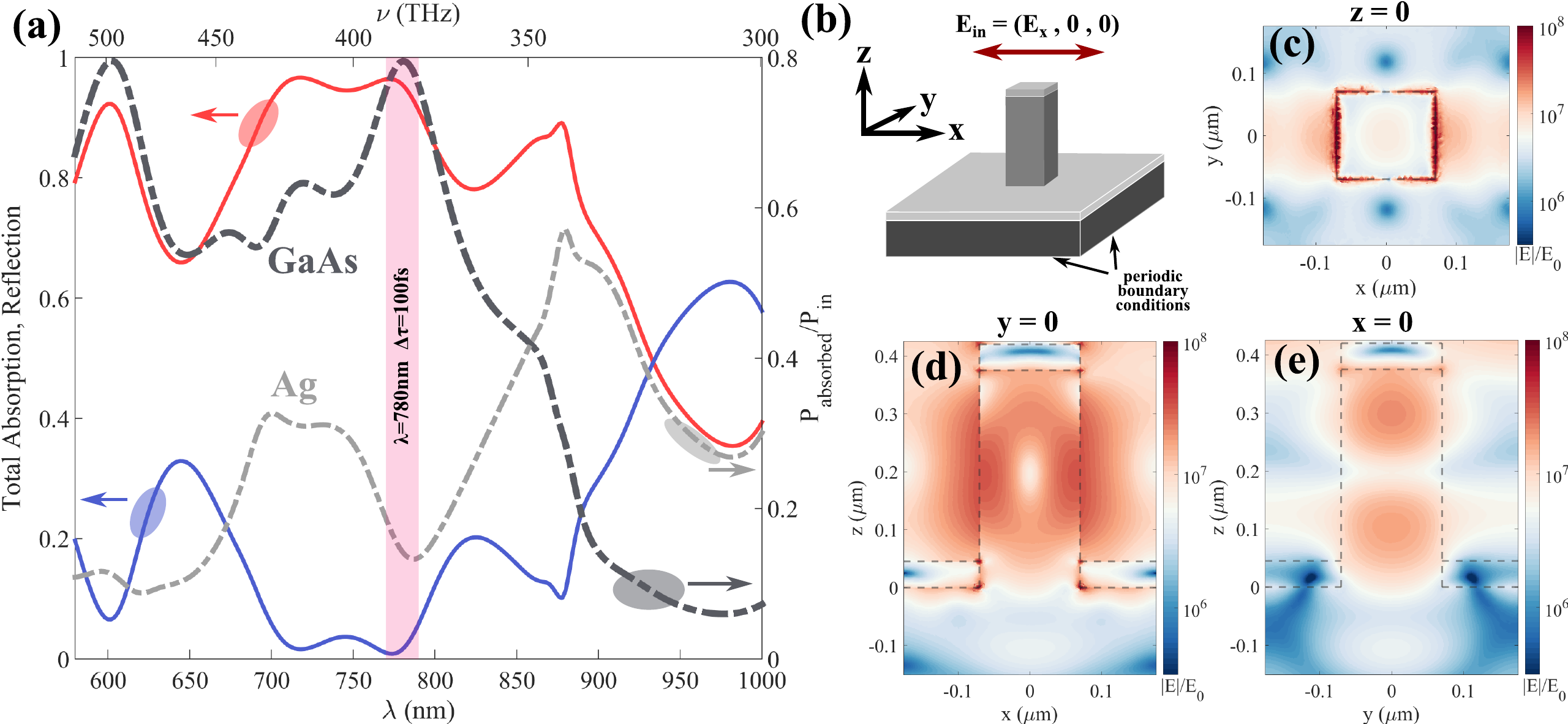}
\caption{\textbf{Spectrum and near-field enhancement.} \textbf{a}. Spectral performance of the nanopillar photo-conductive switch. The left axis shows the total reflection (blue) and absorption (red) by the nanopillar device. The right axis illustrates the overall absorbed power inside the semiconductor (black dashed line) and inside the metal contacts (grey dashed line). The spectral bandwidth of a $100 \ \rm{fs}$ pulsed laser is shown by the red box centred at $\lambda = 780\ \rm{nm}$. 
\textbf{b}. Schematic that represents the excitation and boundary conditions used for our simulations.
\textbf{c-e}. Near field enhancement maps at the excitation wavelength ($\lambda = 780\ \rm{nm}$) and for the optimum geometric conditions discussed in the text. The effect of coupling to a propagating surface plasmon (SPP) wave is shown by \textbf{c}. The coupling to the SPP is weak, but we can see however a strong local field enhancement at the vicinities of the pillar. The coupled TE$_{0}$ waveguide mode and the $m=2$ Fabry-Perot mode are illustrated by \textbf{d,e}. The overall field enhancement in the nanopillar region is almost 2 orders of magnitude higher than elsewhere.}
\label{fig2:spectrum_near_field}
%\end{center}
\end{figure}

As pointed out earlier, to increase the overall performance of the photo-conductive switch we excite photonic and plasmonic modes in the device by positioning the nanopillars in a periodic array and by carefully choosing its height.
As illustrated by the inset of Fig. \ref{fig1:schematic_dispersion}b, the cross section of the proposed device forms a waveguide. The top (ITO) and bottom (Ag) metallic electrodes constitute the claddings of the waveguide, whereas transparent insulator and array of nanopillars form an effective medium which is the core of the waveguide.
The effective refractive index of the insulator/nanopillars medium is defined by the geometric average between the GaAs and the insulator, namely $n_{\rm eff} = 2.38$ and the thickness of the waveguide is equal to the height of the nanopillars, $d_{\rm wg} = h = 330\ \rm{nm}$.
Figure \ref{fig1:schematic_dispersion}b shows in blue colour the dispersion curves for the first two modes of this waveguide. The region on the right of the ITO light cone (black line) defines the bound modes and the grey area on the left defines the leaky modes.
A plane wave incoming from the top of the device, under normal circumstances, cannot excite these modes because of momentum mismatch. 
By placing the nanopillars in a periodic lattice, we can add extra in-plane momentum (x-y direction, $k_{||}$) to the incoming plane wave and thus excite the transverse electric (TE) propagating waveguide modes into the effective medium defined by the GaAs nanopillars and transparent insulator.
We chose our working wavelength to be $\lambda=780\ \rm{nm}$, since commercial femtosecond lasers are readily available at this wavelength, such as Ti:Sapphire or doped fibre lasers. At this wavelength, and with the periodicity of our device being $P=355\ \rm{nm}$ we can excite the TE$_0$ waveguide mode, shown by the large red diamond.
While the TE$_0$ mode travels through the waveguide it couples to a Fabry-Perot mode that exists along the height ($h=330\ \rm{nm}$) direction of the nanopillar. The second order Fabry-Perot mode, shown by the grey dashed line, is a dark mode and it cannot be excited by the incoming plane wave due to the presence of the nanopillar top metallic contact (Ag). It can be excited however through its coupling to the TE$_0$ waveguide mode.
In addition to the two photonic modes supported by our device a third, plasmonic mode, exists.
At the laser wavelength and for a nanopillar width of $w=140\ \rm{nm}$, our periodic structure can weakly couple to a propagating surface plasmon polariton (SPP)\cite{ebbesen1998,abajo2007}. The SPP is excited at the interface between the semiconductor substrate and the bottom metal electrode. The coupling to the SPP is weak since its excitation wavelength is at $\lambda = 885\ \rm{nm}$ (see Fig. \ref{fig2:spectrum_near_field}a). 
As we will elucidate further in the text, even though the coupling to a SPP is  weak, our device profits from large local field enhancements at the vicinities of the nanopillars.

Since the proposed photo-switch can support three types of photonic/plasmonic modes that are different in nature, it is expected that it can absorb completely the input laser beam. Our simulations show that this is indeed the case, as illustrated by Fig. \ref{fig2:spectrum_near_field}a. The red and blue continuous lines correspond to the overall absorption and reflection respectively (see left axis), when the nanopillars are illuminated by a plane wave at normal incidence. The light coloured red box centred at $\lambda=780\ \rm{nm}$ represents the spectral bandwidth of a $\Delta\tau=100\ \rm{fs}$ laser pulse assuming that it is Fourier transform limited. At this wavelength the laser beam is completely absorbed ($>$ 98\%) by the device, while the reflectance is reduced to almost zero percent. 
The zero back-reflectance from the top surface of our sample, is also a consequence of optimising the thickness of the transparent conducting electrode (ITO). By choosing the right thickness we can form an anti-reflection coating at the surface of our device, while at the same time have a conductive electrode for collecting the photo-excited carriers. This is achieved at an ITO thickness of $h_{\rm{ITO}} = 125 \rm{nm}$. The right axis of Fig. \ref{fig2:spectrum_near_field}a, shows the absorbed power normalised to the incident power in both the semiconductor region (black dashed line) and in the metallic contacts (light grey dashed line). The absorbed power is calculated by integrating in the semiconductor or metal (Ag) region the optical power absorbed per unit volume,
\begin{equation}
    P_{\rm{abs}} = -\frac{1}{2} \omega \epsilon_0 \int_{\rm{V}} dV |E(x,y,z)|^2 \ \Im \{ \epsilon(x,y,z) \}
    \label{eq:absorbed_power}
\end{equation}
where $\epsilon_0$ is the vacuum permittivity, $\omega$ is the angular frequency,  $E(x,y,z)$ is the electric field and $\Im \{ \epsilon \}$ is the imaginary part of the permittivity in each material. As we can appreciate from these data, although the overall absorbed power is almost unity, the input energy is not exclusively absorbed by the semiconductor. A ratio of around 20\% is absorbed by the metal contacts and is potentially converted into heat through resistive losses. We note that resistive losses in the metal contacts can significantly deteriorate the efficiency of photo-conductive switches, especially those who rely solely on plasmonic resonances.

%%%%%%%%%%%%%%%%%%%%%%%%% Fig. 2 %%%%%%%%%%%%%%
%%%%%%%%%%%%%%%%%%%%%%%%% Fig. 2 %%%%%%%%%%%%%%

For our simulations, to reduce metal losses as much as possible, we have compared several metals with various permittivities. The metals with the best performance for our device were gold (Au) and silver (Ag), with Ag being slightly better than Au. This is because losses in Ag are lower than Au at $\lambda=780\ \rm{nm}$ and this is reflected by their permittivity values, namely $\epsilon = -29 + 0.37i$ for Ag and $\epsilon = -22 + 1.4i$ for Au. The optimum thickness for the metal electrodes covering the bottom surface and the top part of each nanopillar was found to be $h_{\rm{m}} = 45 \ \rm{nm}$. 

Figures \ref{fig2:spectrum_near_field}(c-e) show in logarithmic scale the near field enhancement distribution of the vertical photo-conductive switch at $\lambda=780\ \rm{nm}$ and for the optimum parameters discussed above. From the x-z and y-z cross-sections, at $y = 0 \ \rm{\mu m}$ and $x = 0 \ \rm{\mu m}$ respectively (Fig. \ref{fig2:spectrum_near_field}(d-e)), it is evident that the majority of the electric field is concentrated in the semiconductor region. The field enhancement in the nano-pilar region due to the coupling between the TE$_{0}$ waveguide mode and the $m=2$ Fabry-Perot mode reaches levels that are almost 2 orders of magnitude higher than elsewhere. In addition, we can see from Fig. \ref{fig2:spectrum_near_field}c that at $\lambda = 780\ \rm{nm}$ our structure is weakly coupled to a SPP, propagating at the interface between the metal (Ag) and the substrate ($z = 0\ \rm{\mu m}$). The SPP, normally excited at $\lambda = 885\ \rm{nm}$, has a broad linewidth (see Fig. \ref{fig2:spectrum_near_field}a) which extends to the operating wavelength of our device. The coupling to a SPP, although weak, causes a significant electric field enhancement at the vicinity of the nanopillar\cite{ebbesen1998}. 

\subsection*{Photo-conductive switch semiconductor modelling}
The electronic properties of a single unit cell of our device are numerically simulated with the Poisson-Drift-Diffusion equation. This is done by using the same commercial package as the one used for modelling the photonic response of our device (see Methods section).
The main components of our model are the electron generation term, the recombination rate component and a carrier mobility model.

To account for the complicated carrier dynamics in our nanopillar device, the semiconductor equations are solved with a time-domain solver.
The electron generation term is derived directly from the photonic simulation and, more precisely, from the optical absorption per unit volume, shown in Eq. \ref{eq:absorbed_power}. This term is then multiplied by a Gaussian time-dependent function to account for the laser excitation, resulting,
\begin{equation}
    G(x,y,z,t) = \frac{P^2}{\pi r^2} \ \frac{\mathcal{P}_{av}}{ f_{L} \sqrt{\pi} \Delta\tau} \ \frac{\mathcal{P}_{abs}(x,y,z)}{E_{\nu}} \ e^{-\frac{(t-t_0)^2}{\Delta\tau^2}}
    \label{eq:generation_term}
\end{equation}
where the first term of the formula defines the fraction of the illuminated unit cell area with respect to the area of the laser beam, with  $P$ being the periodicity of our device and $r$ the radius of the laser beam.
The second part of the equation defines the peak power delivered by a femtosecond laser with $\Delta\tau$ pulse width, $f_{L}$ repetition rate, and an average power of $\mathcal{P}_{av}$. 
The third part of the generation term accounts for the spatial absorption within the semiconductor structure, with the assumption that one photon of energy $E_{\nu} = hc/\lambda$ will generate one electron-hole pair. 
The assumption of the quantum efficiency being equal to unity is a simplification to our model since taking into account any quantum effects will result into numerically cumbersome calculations.
Finally, the last term in the generation formula is a Gaussian function to account for the temporal laser pulse shape, which arrives at time $t_0$.

For the carrier recombination term, we used the well established trap assisted Shockley-Read-Hall model.
This model takes into account an ensemble of carrier traps that are formed in the LTG-GaAs during epitaxial growth and subsequent annealing with an effective recombination time, $\tau_{\rm r} = 0.8\ \rm{ps}$\cite{Gupta1991}. 
It is expected that the carrier recombination as well as the mobility will not be significantly affected by any surface effects, as the proposed nanopillars are bigger than previously studied nano-wire devices\cite{Joyce2017}. In addition, previous work on nano-patterned THz photo-conductive switches suggests that there is no degradation of the LTG-GaAs electrical properties\cite{siday2019}. As a consequence, for our simulations, we consider that the electrical properties of the nanopillars are obtained from the properties of bulk LTG-GaAs. For simplicity, we have also assumed that the trapping rates for electrons are the same as for holes.

%%%%%%%%%%%%%%%%%%%%%%%%% Fig. 3 %%%%%%%%%%%%%%
\begin{figure}[t!]
%\begin{center}
\includegraphics[width=0.95\linewidth]{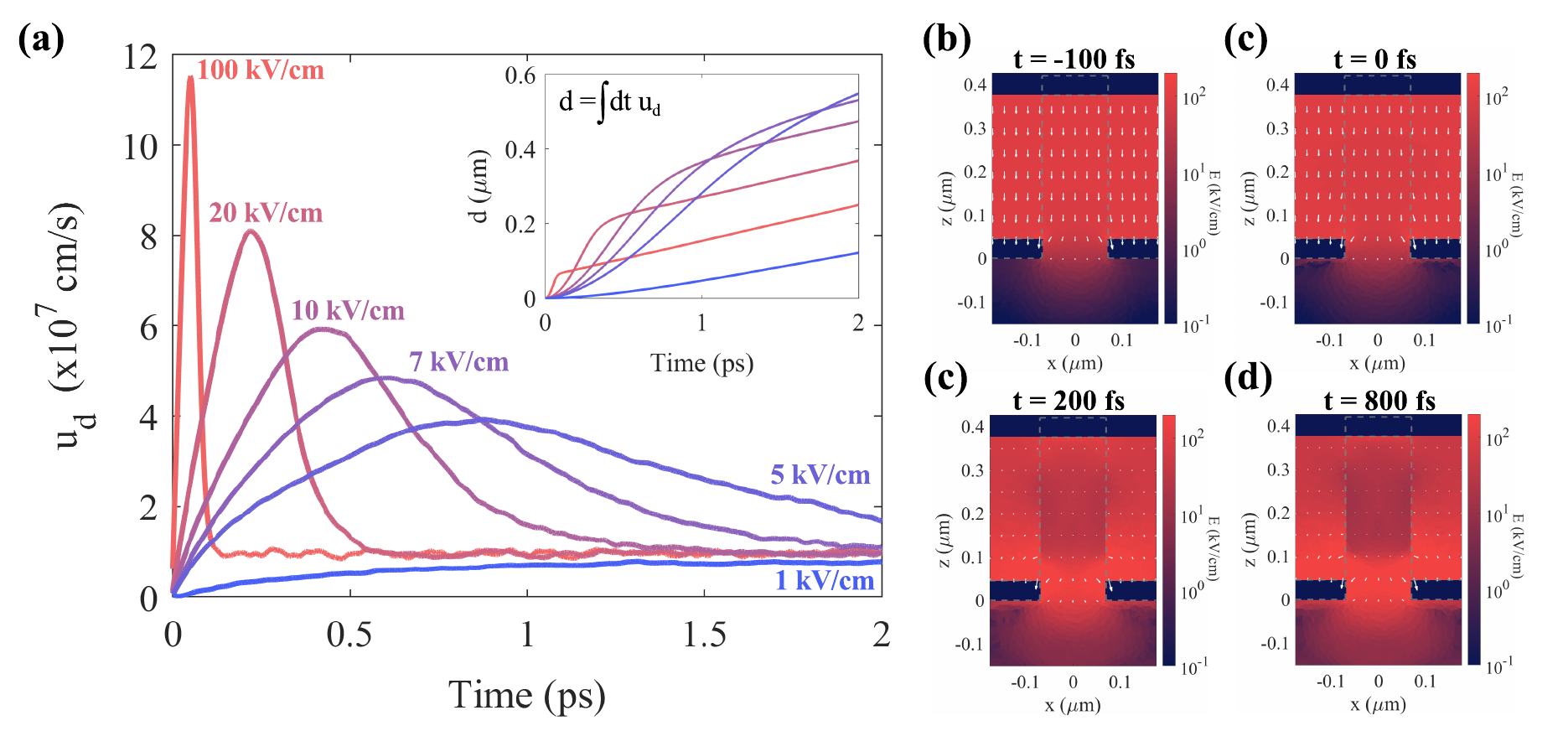}
\caption{\textbf{Drift velocity and electric potential screening.} \textbf{(a)} Electron drift velocity as a function of time for electric fields in the range of 1 $\rm{kV/cm}$ to 100 $\rm{kV/cm}$. The velocity is calculated using a Monte-Carlo simulator with electron density $N=10^{18} \ \rm{cm^{-3}}$. 
The inset shows the average distance travelled by the electrons. \textbf{(b--e)} Electric potential distribution in the nanopillar photo-conductive switch for a $3\ \rm{V}$ bias voltage, corresponding to an electric potential of about $90\ \rm{kV/cm}$.
\textbf{(b)} Before the arrival of the optical pulse, $t = -100\ \rm{fs}$ \textbf{(c)} Arrival of the optical pulse, $t = 0\ \rm{fs}$ 
and \textbf{(d,e)} Electric potentials at times $t = 200\ \rm{fs}$ and $t = 800\ \rm{fs}$ after the carriers are photo-excited. The white arrows illustrate the direction of the electric potential.}
\label{fig4:drift_velocity}
%\end{center}
\end{figure}
%%%%%%%%%%%%%%%%%%%%%%%%% Fig. 3 %%%%%%%%%%%%%%

In contrast to the ``classical'' interdigitated photo-conductive switch design, the photo-excited carriers in our vertical structure can be accelerated towards the top and bottom electrodes by very high electric fields. For instance, a bias voltage of $5\ \rm{V}$ can induce an electric field in the order of $150\ \rm{kV/cm}$. 
Electron and hole drift velocities, and as a consequence carrier mobilities, are largely influenced by these high fields. 
This is because carriers acquire very large kinetic energies that lead to a non-equilibrium dynamic response in the picosecond and sub-picosecond time scale. To account for the out-of-equilibrium carrier dynamics we have used a Monte-Carlo model to simulate the electrons' drift velocity and estimate the carrier mobility as a function of time. 
Our Monte-Carlo simulations are based on an open source simulator\cite{vasileskaMC}, in which we have used a carrier density of $N = 10^{18} \ \rm{cm^{-3}}$ and $10^4$ events per simulation. 
This carrier density reflects the actual amount of electrons and holes generated in our device by a femtosecond laser pulse with an average power $\mathcal{P}_{\rm av} \sim 1\ \rm{mW}$, pulse duration $\Delta\tau= 100\ \rm{fs}$, beam radius $r=20\ \rm{\mu m}$ and a repetition rate of $f_{L}=100\ \rm{MHz}$. 
The model takes into account the $\Gamma$, $X$, and $L$ valleys of GaAs with energies $1.42\ \rm{eV}$, $1.71\ \rm{eV}$, $1.90\ \rm{eV}$ and effective masses $0.063\rm{m_e}$, $0.87\rm{m_e}$, $1.98\rm{m_e}$ respectively.
Our results are illustrated in Fig. \ref{fig4:drift_velocity}, which shows the carrier drift velocity as a function of time for different electric fields. At relatively low electric fields, $\sim 1\ \rm{kV/cm}$ (blue curve), the carriers accelerate slowly for the first $1-1.5 \ \rm{ps}$ until they reach their saturation velocity, $u_{\rm{sat}} \sim 10^7 \ \rm{cm/s}$.
At moderate electric fields, $5-10\ \rm{kV/cm}$, consistent with what has been reported in the past\cite{Pozela1980,Foutz1999,Reklaitis1999}, we observe a velocity overshoot up to five times the saturation velocity, which lasts for almost $0.8\ \rm{ps}$. This sudden increase of the drift velocity is attributed to the very high kinetic energies attained by the carriers. At short times $<0.5 \ \rm{ps}$, electrons accelerate and gain speed in the $\Gamma$ valley of the conduction band where the effective mass is the smallest. 
However, at longer times the carrier energy exceeds the threshold for enabling transitions to the $X$ and $L$ bands which are higher in energy. 
As the electron effective mass in these bands is much bigger than the lower energy band, the electrons will begin to loose speed, eventually reaching a saturation velocity that is dominated by scattering mechanisms with the lattice.
Finally, at very high electric fields the overshoot velocity can reach values up to $10^8 \ \rm{cm/s}$ within the first $100 \ \rm{fs}$.
The inset of Fig. \ref{fig4:drift_velocity}, shows the integral of the drift velocity, which corresponds to the average distance travelled by the electrons. As we can see from the data, at very high electric fields, although the carriers achieve high overshoot velocity, since it only lasts for a few femtoseconds, they will not travel very far. 
For our 3-dimensional nanopillar photo-conductive switch, the electrons would have to travel an average distance of $0.2-0.25 \ \rm{\mu m}$ towards the contacts in less than $1\ \rm{ps}$ (recombination time). There is indeed a clear trade-off between the applied bias electric potential and the amount of carriers that can make it to the collection electrodes before they are captured by the traps.

As pointed out earlier, the vertical design of our photo-conductive switch has the advantage of inducing homogeneous electric potentials when a bias voltage is applied between the top and bottom electrodes. 
This is in stark contrast to the existing planar photo-conductive switches, where the electric potential becomes very weak in the bulk of the semiconducting LTG-GaAs. 
Positioning the two electrodes on the opposite ends of the photo-active region, in a similar way to a capacitor, can significantly increase the overall power performance of the photo-conductive switch\cite{Peytavit2009,Peytavit2011a,Peytavit2011b}. 
It can, however, as we will discuss later, influence the electrical properties of the generated pulses.

Figures \ref{fig4:drift_velocity}(b--e) show the electric potential distribution for our nanopillar photo-conductive switch for a bias voltage of $3\ \rm{V}$. This bias voltage corresponds to an electric potential of about $90\ \rm{kV/cm}$. At negative times, before the arrival of the optical pulse ($t = -100\ \rm{fs}$), the electric potential lines are homogeneous and of equal amplitude within the nanopillar region. A very similar situation is shown at $t = 0\ \rm{fs}$, the time when the femtosecond pulse photo-excites the electron hole pairs in the LTG-GaAs. There are a few subtle differences, however the intensity and the direction of the electric potential lines remain the same.
The homogeneous electric potential picture, on the other hand, completely changes at longer times, i.e. $t = 200\ \rm{fs}$ and $t = 800\ \rm{fs}$. The intensity of the electric potential drops significantly in the nanopillar region by a factor of two at $t = 200\ \rm{fs}$ and by almost one order of magnitude at $t = 800\ \rm{fs}$. This means that the driving force of the electron-hole pairs is reduced from $90\ \rm{kV/cm}$ to  $15\ \rm{kV/cm}$ within $800\ \rm{fs}$. 
This sudden drop of the electric potential is caused by the $N \sim 10^{18} \ \rm{cm^{-3}}$ photo-generated carriers that strongly screen the electric potential in regions where the carrier density is the highest. 
This, however, does not limit the performance of our 3-dimensional design. As will be elaborated in the next paragraphs, the ultrafast response of the nanopillar photo-conductive switch is primarily dominated by the mobility of the photo-generated carriers.

%%%%%%%%%%%%%%%%%%%%%%%%% Fig. 4 %%%%%%%%%%%%%%
\begin{figure}[t!]
%\begin{center}
\includegraphics[width=0.8\linewidth]{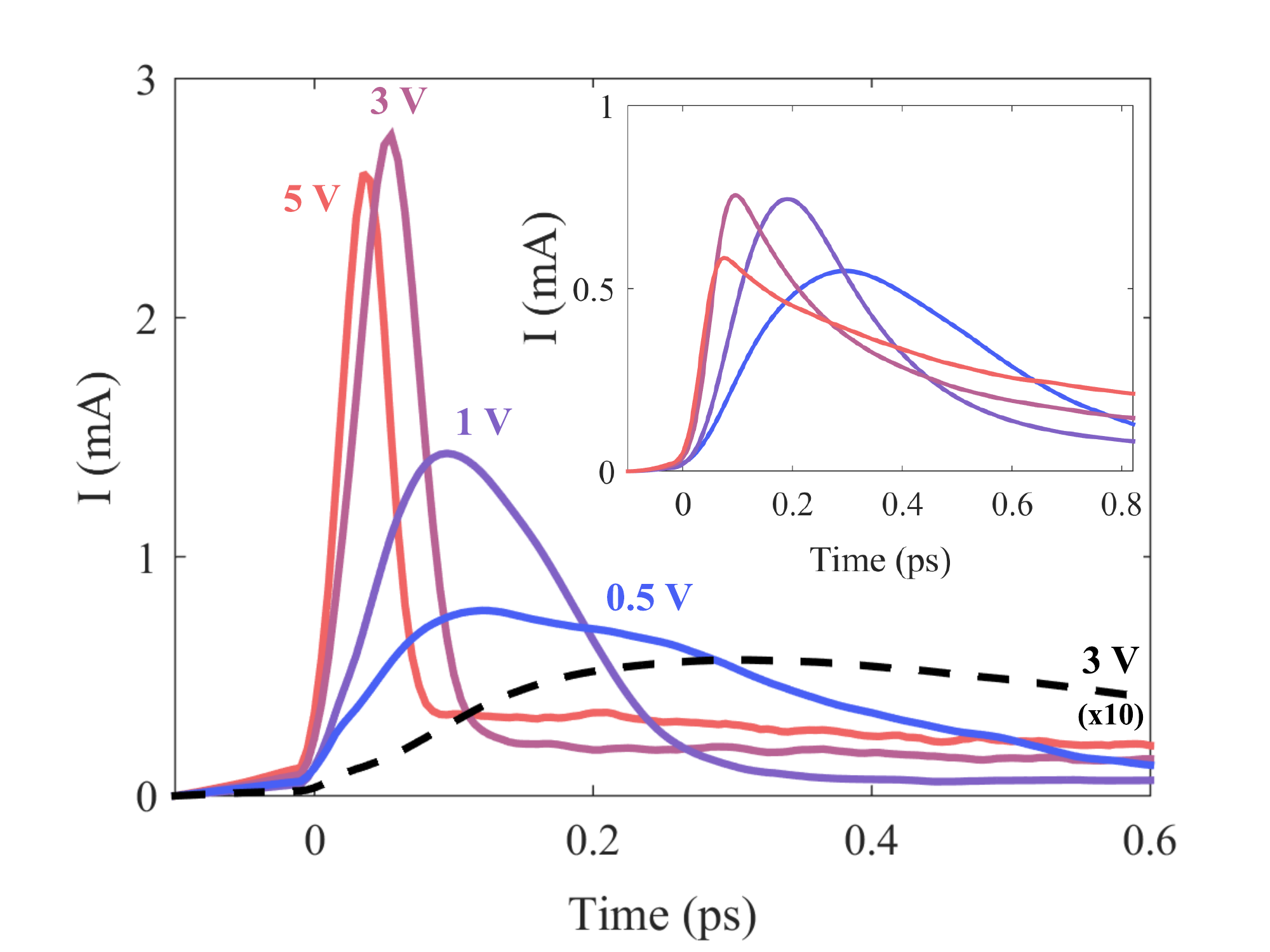}
\caption{\textbf{Photo-generated current.} Generated current for an input laser power of $1\ \rm{mW}$ and for a single nanopillar unit cell. The coloured lines show the current for various bias voltage, namely from $0.5\ \rm{V}$ to $5\ \rm{V}$ corresponding to electric fields of $15\ \rm{kV/cm}$ and $150\ \rm{kV/cm}$ respectively. The dark dashed line shows the same simulation for a ``classical'' interdigitated Auston photo-conductive switch with a bias votage of $3\ \rm{V}$. The simulation data for the ``classical'' photo-conductive switch are multiplied by a factor of 10 to illustrate the differences. The inset shows the outcome of applying a low-pass filter to the photo-generated pulses with $\tau_{_{\rm{RC}}} \sim 1\ \rm{ps}$.}
\label{fig5:generated_current}
%\end{center}
\end{figure}
%%%%%%%%%%%%%%%%%%%%%%%%% Fig. 4 %%%%%%%%%%%%%%

To quantify the overall photo-current generated by each nanopillar we have numerically calculated the current flowing through the top and bottom electrodes of our device. 
The time-dependent carrier mobility model used for these simulations takes into account the possible velocity overshoot demonstrated by our Monte-Carlo simulations.
Figure \ref{fig5:generated_current} shows the generated current due to the photo-excited electrons for a range of bias voltage, ranging from $0.5\ \rm{V}$ to $5\ \rm{V}$. 
For $0.5\ \rm{V}$ (blue curve), we observe a pulse with a current amplitude of $800\ \rm{\mu A}$ and a time width of $\sim 300\ \rm{fs}$. 
For larger bias voltage, $3\ \rm{V}$, the generated pulse has an amplitude that reaches values as high as $3\ \rm{m A}$ per nanopillar. 
This very high amplitude pulse is also linked with a very fast dynamic response, of about $\sim 100\ \rm{fs}$, which is in turn accompanied by a slower dynamic response at times $t > 100\ \rm{fs}$ related to carrier recombination. 
The largest peak current amplitude is obtained for a bias voltage of $3\ \rm{V}$, whereas for higher bias voltage we observe a small reduction of the current peak. 
This is associated with the drift velocity of the carriers and the distance they can travel before reaching their saturation velocity (see inset of Fig. \ref{fig4:drift_velocity}a).

To benchmark the efficiency of our novel photo-conductive switch, we compare our results to a ``classical'' interdigitated Auston photo-conductive switch, shown with a black dashed line on Fig. \ref{fig5:generated_current}.
For this simulation, we have considered an interdigitated photo-conductive with a $2\ \rm{\mu m}$ electrode width and a $5\ \rm{\mu m}$ periodicity. This ``classical'' device had LTG-GaAs thickness of $1\ \rm{\mu m}$, while the carrier recombination time and mobility was kept the same as the model used for the nanopillar device.
To keep the comparison on an equal footing we have normalised the results with the surface area of the nanopillar photo-conductive switch.
We observe that for a bias voltage of $3\ \rm{V}$, our 3-dimensional device outperforms the ``classical'' interdigitated switch by a factor of 50 in peak current amplitude. 
In addition, for the planar design, the pulse width of the generated pulse is limited to around $800\ \rm{fs}$. 
This is because the electric potential for this 2-dimensional planar structure does not drive the free carriers as strongly as it does for the nanopillars and this results in the recombination of the free carriers before they reach to the collection electrodes.

\subsection*{Discussion}
The results of Fig. \ref{fig5:generated_current} suggest that the proposed 3-dimensional nanopillar design can outperform ``classical'' planar designs by a factor of 50 in terms of the generated current amplitude. 
The absolute amplitude of the photo-generated current can be linearly increased with the number of nanopillars within the device. 
Although in our photonic simulations we are simulating a periodic array of nanopillars, which means an infinitively long photo-conductive switch, for realistic applications the device has to be small enough to be integrated in any nano-electronic circuit. 
In addition, the overall footprint of the device adds some extra constraints that can limit its efficiency.

To begin with, it is important to note that the proposed device cannot be made of a single nanopillar.
This is because, for the collective photonic and plasmonic effects to take place, a minimum of fifteen periods should be used\cite{Rodriguez2012}.
Efficient excitation of all the optical modes supported by our device is important to achieve maximum performance.
This sets a lower limit to the device area at about $5.5\ \rm{\mu m} \times 5.5\ \rm{\mu m}$.

In addition, as we have seen from Fig. \ref{fig5:generated_current}, the generated pulse can have a time response as short as $100\ \rm{fs}$ for very high acceleration fields. 
Even though very high bandwidth THz pulses have been experimentally observed in the past for very high acceleration fields\cite{Shen2003, Dreyhaupt2005}, the area of the device can significantly limit its bandwidth. 
This is because opto-electronic devices, such as the one presented here, have a characteristic internal time constant $\tau_{_{\rm RC}} = RC$, which practically renders them into low pass filters. 
The cut-off frequency, $f_{\rm c} = 1/\tau_{_{\rm RC}}$, of the photo-conductive switch is inversely proportional to the total capacitance of the device.
To estimate the capacitance of the proposed photo-conductive switch we assume a parallel plate capacitor. The two parallel electrodes for this capacitor are the top and bottom metallic contacts of the photo-conductive switch and the dielectric between the contacts is composed of the GaAs nanopillar ($\epsilon = 12.8$) and the transparent insulator surrounding the nanopillar ($\epsilon = 1.58$). Given that each nanopillar unit-cell has a typical capacitance of $C_{\rm np} = 15\ \rm{aF}$ and that all nanopillars are connected in parallel, the total capacitance for a 15 by 15 array would be $C_{\rm tot} = 3.5\ \rm{fF}$. This sets the time constant of the photo-conductive switch at $\tau_{_{\rm RC}} = 350\ \rm{fs}$, assuming a $100\ \rm{\Omega}$ load resistance\cite{Brown1993}.
The inset of Fig \ref{fig5:generated_current}, presents an extreme situation, where the internal time constant of the photo-switch is set at $\tau_{_{\rm RC}} = 1\ \rm{ps}$, which is translated into an 26 by 26 nanopillar array. 
It is clear that not only the time-width of the pulse is increased to almost $1\ \rm{ps}$ but also the peak current is reduced by a factor of four. Even for such extreme cases, the performance of our 3D device is still much larger than the classical planar device.

\subsection*{Conclusion}
In conclusion, we have presented a novel photo-conductive switch device that has a 3-dimensional design. The optical to electrical power conversion efficiency of the proposed device is significantly increased by maximising the optical absorption in the LTG-GaAs through photonic and plasmonic resonances. In addition, by positioning the electrodes of the device at the end-sides of the LTG-GaAs we can uniformly accelerate the electrons under high electric potentials, thus reaching high carrier collection efficiencies. 
Our numerical results suggest that our nanopillar device can deliver electrical pulses with as much as 50 times more peak current and significantly broader bandwidth than the ``classical'' interdigitated photo-conductive switches. 
These qualities, will be useful and attractive to quantum information experiments as well as applications that require ultrafast voltage pulses at cryogenic temperatures\cite{dubois2013,lin2018,roussely2018}.

\subsection*{Methods}
The 3-dimensional photonic-plasmonic photo-conductive switch was modelled using the commercial software Comsol Multiphysics v5.2a. 
For our simulations we have constructed a coupled electromagnetic (RF) and semiconductor (Semi) model, that first solves the electromagnetic problem whose solutions are then fed as an input to the semiconductor problem. 
As the photo-conductive switch is made of a periodic array of nanopillars, we had to solve the RF and Semi models only for a single unit cell. A schematic of the unit cell is shown in Fig. \ref{fig1:schematic_dispersion}a. The substrate and nanopillar are composed of LTG-GaAs and the metal electrodes (top and bottom) are Ag with $45\ \rm{nm}$ thickness. 
The nanopillar has a height of $h=330\ \rm{nm}$ and is surrounded by a transparent insulator/dielectric with a refractive index of $n=1.58$. The width of the nanopillar is $w=140\ \rm{nm}$ and the periodicity of the array is $P=355\ \rm{nm}$. To promote anti-reflection properties at the excitation wavelength $\lambda=780\ \rm{nm}$, we place on the top part of the device a transparent electrode (ITO) with thickness, $h_{\rm{ITO}} = 125\ \rm{nm}$.
We used periodic boundary conditions for the electromagnetic model, in both x and y directions. A plane wave is launched from the top part of the simulation region (+z) and propagates through air ($n=1$) towards the bottom of the device (-z). The input optical power for all of our simulations was $1 \rm{mW}$ defined for an optical beam with a diameter of $20\ \rm{\mu m}$ and appropriately adjusted for the $355\rm{nm}\times 355\rm{nm} $ area of the nanopillar unit cell. The transmission, reflection and absorption data shown in Fig. \ref{fig2:spectrum_near_field}a are extracted directly from the S-parameters provided by the simulation, whereas the absorbed power inside the nanopillar and Ag are calculated using Eq. \ref{eq:absorbed_power}.

To calculate the total photo-generated current we solved the time-dependent semiconductor model only in the LTG-GaAs regions, that is the nanopillar and substrate. Insulation boundary conditions were used for this model on all x-y sides, and Ohmic contacts with $V=V_{\rm{bias}}$ and $V=0\ \rm{Volts}$ for the top and bottom electrodes respectively. To account for the spatial distribution of the photo-generated carriers within our device we used an electron/hole generation term that takes into consideration the results calculated from the electromagnetic calculation and the laser parameters (see Eq. \ref{eq:generation_term}). The recombination time used in our simulations was $\tau=0.8\  \rm{ps}$ for both electrons and holes. The mobility of the photo-generated carriers is derived from a Monte-Carlo (MC) simulation model as described in section ``Photo-conductive switch semiconductor modelling''. From the MC simulation we have calculated the drift velocity of carriers ($u_{\rm{d}(t,E)} = \mu(t,E)\ E$) as a function of time, $t$, and applied electric potential, $E$, (see Fig. \ref{fig4:drift_velocity}). To simplify our simulations we have assumed that for the mobility calculation, the electric potentials is given by $E=V_{\rm{bias}}/h$, where $V_{\rm{bias}}$ is the bias voltage applied to the top electrode and $h$ is the height of the nanopillar. This assumption, is to a first approximation correct as the two electrodes are positioned parallel to each other and the electric field distribution is homogeneous within the nanopillar region.

%===========================%===========================%=====================
\newpage
\section*{Acknowledgments}
The authors would like to thank Jean-Louis Coutaz for fruitful discussions throughout the duration of this project. 
This work was supported by the ANR QTERA project, grant ANR-2015-CE24-0007-02 of the French Agence Nationale de la Recherche.

%%%%%%%%%%%%%%%%%%%%%%%%%%%%%%%%%%%%%%%%%%%%%
%%%%%%%%%%%%%% Bibliography %%%%%%%%%%%%%%%%%
%%%%%%%%%%%%%%%%%%%%%%%%%%%%%%%%%%%%%%%%%%%%%
%\begin{thebibliography}{10}
\bibliography{refs}
%\end{thebibliography}

%\section*{Author contribution}

%\textbf{Competing financial interests:} The authors declare no competing financial interests.

%\end{linenumbers}
\end{document}